\documentstyle[aps,preprint,12pt]{revtex} 

\begin{document}

\newcommand{\bel}{\begin{equation}\label}
\newcommand{\f}{\frac}
\newcommand{\bee}{\begin{equation}}
\newcommand{\ee}{\end{equation}}
\newcommand{\br}{\begin{eqnarray}}
\newcommand{\brr}{\begin{eqnarray*}}
\newcommand{\er}{\end{eqnarray}}
\newcommand{\err}{\end{eqnarray*}}
\newcommand{\pr}{\partial}
\newcommand{\non}{\nonumber \\}
\def\tr{\hbox{tr}} 
\def\th{\theta} 
\renewcommand{\thesection}{\arabic{section}}
\renewcommand{\thesubsection}{\thesection.\arabic{subsection}}
\renewcommand{\theequation}{\thesubsection.\arabic{equation}}

\title{
\hfill\parbox{4cm}{\normalsize IMSc/98/12/58 \\hep-th/9902087}\\ 
\vspace{2cm}High Temperature Limit of the $ N= 2 $  Matrix Model}
\author{Subrata Bal and  B. Sathiapalan 
\footnote{ email - subrata,bala@imsc.ernet.in}}
\address{The Institute of Mathematical Sciences, 
 CIT Campus, 
Madras - 600 113,  INDIA}
\maketitle

\begin{abstract}

	The high temperature limit of a system of two D-0 branes is 
investigated. The partition function can be expressed as a power series 
in $\beta$ (inverse temperature). The leading term in the high temperature 
expression of the partition function and effective potential is 
calculated {\em exactly}. Physical quantities like the mean square 
separation can also be exactly determined in the high temperature limit.

\end{abstract}

\section{Introduction}
\setcounter{equation}{0}

The study of string theory at finite temperature 
 has received renewed attention recently 
\cite{BS1,KRBS,MOP,MP,AMB,CM,TSU,MINC,LT,POL,MG,WEX,KOSV,TSEY,MAVM,HAL}.
In a recent paper one of us \cite{BS1} attempted to elucidate
the nature of the Hagedorn transition \cite{H}  using the matrix model and found
 similarities with the deconfinement transition in gauge theories.
This was also investigated in a subsequent paper using the AdS/CFT
correspondence \cite{KRBS}. It is clear that recent developments in
 non-perturbative string theory or M-theory 
\cite{WIT,BFSS,IKKT,FKKT,AIKT} have some bearing
on our understanding of the high temperature behavior of strings.
 Furthermore, study of high temperature behavior of a
 system is often a useful probe of the system  because many
 calculations  simplify at high temperature. We can thus hope to
 learn something about M-theory from its high temperature behavior.
With this motivation, in this paper, we attempt  to study the high
 temperature
behavior of a simple but
 non-trivial system - the system of two D-0-branes.  
This is essentially the BFSS matrix model \cite{BFSS} with N=2. This is
 not
big enough to describe M-theory. In particular it would not include
for instances processes involving pair-production of 
D-0 brane - anti-D-0 brane. 
Nevertheless it is already
 complicated
enough.  In particular the nature of the threshold and other bound
 states that have 
been studied \cite{PY,SS,SUN,DKPS,MLI} are not fully understood.
Furthermore we should keep in mind that while the matrix model 
reproduces string theory at 
short distances, the fact that it also does so at long distances seems to be 
entirely due to the super-symmetric non-renormalization theorems. 
At finite temperature supersymmetry is broken and perhaps
we should not expect this. For all these reasons the study of matrix models
 at high temperature is worthwhile.

A related model of D-instantons,
the IKKT matrix model \cite{IKKT}, which is 0+0 dimensional 
has been solved exactly for N=2 \cite{TSU}. The D-0-brane action that we are
interested in, is a quantum
mechanical one (i.e. 0+1 dimensional). However after compactifying
the Euclideanised time, if one takes the high temperature limit, it
reduces to a 0+0 dimensional model. There is thus a hope of solving this model
 order by order in $\beta$ but {\em to all orders in $g_s$} using the
 same techniques as \cite{TSU}. One can
then calculate physical quantities such as the mean square
separation of the D-0-branes - a measure of the size of the bound
states.  This is what is attempted in this paper.
We obtain the leading behaviour in $\beta$.
We can also estimate, 
the corrections to the leading result. The noteworthy feature being
that
each term is 
{\em exact in its dependence on the string coupling constant }.

This paper is organized as follows.
In section \ref{ac}  we set up the problem and in section \ref{partf}  we describe
briefly the solution. The last section contains a summary
of the result and some conclusions.

\section{The action}\label{ac} 
\setcounter{equation}{0}

\noindent
The $(0+1)$ dimensional BFSS lagrangian is
\bel{lag1}
L={1\over 2g l_s} \tr \left[  \dot{ X^i}\dot{ X^i} + 2 i \th
 \dot{\th} - {1\over 2 l_s^4}[X^\mu,X^\nu]^2 -
\f{2}{l_s^2} \bar{\th} \gamma_\mu [\th ,X^\mu ] 
- \f{i}{l_s^2} [ X^0, X^i] \dot{X}^i \right]
\ee
where $i = 1,...,9; \mu = 0,..., 9$
; $X^\mu$ and $\th$ are $N $ x  $ N$ hermitian matrices.
$X^\mu$ is 10 a dimensional vector and $\th$ is 16 component Majorana-Wyel 
spinor in 10 dimensional super-Yang Mills theory.
For $N = 2$, we can re-write
$X^\mu$ and $\th$  as 
\bee
X^\mu = \sum_{a =1}^3 {1 \over 2} \sigma^a X^\mu_a,
~~~~~~~~
\th =\sum_{a=1}^3 {1 \over 2} \sigma^a \th_a 
\ee
where,
$\sigma^a$ are the hermitian Pauli matrices and  $X^\mu_a, \th_a $ 
are real fields in terms of which , the Lagrangian (\ref{lag1}) is

\br  
L&=&{1\over 4g l_s} \left\{ 
\sum_{a,b =1}^3
\dot{X^i}_a \dot{X}^i_a
+ 
2i \sum_{a=1}^3 \th_a \dot{\th_a}
-\f{2i}{l_s^2}
\sum_{a,b,c=1}^3 \th_a \gamma_0 \gamma_\mu\th_b X^\mu_c
 \epsilon^{abc}
\right.
\non&+&
\f{ \epsilon^{abc}}{4 l_s^2}
X_a^0 X_b^i \dot{X}_c^i
\left.
+{1\over 2 l_s^4} 
\sum_{a,b=1}^3
X^\mu_a X^\nu_b X^\mu_a X^\nu_b
- 
{1\over 2 l_s^4} 
\sum_{a,b=1}^3
X^\mu_a X^\nu_b X^\mu_b X^\nu_a
\right\}
\er  

\noindent
If we Euclideanize and compactify time on a circle of circumference  $\beta$
the action becomes 

\bee  
S = i \int_0^\beta L dt 
\ee  

\noindent
and the fields satisfy the boundary conditions

\[ X^\mu(0) =  X^\mu(\beta) , ~~~~~~
 \th(0) = - \th(\beta) \]

\noindent
Considering these boundary condition, we can expand the fields $X^\mu, \th $
in modes as

\bee
 X^\mu_a(t) = \sum_{n = - \infty}^{\infty} 
 X^\mu_{a,n} e^{\f{2 \pi i n}{\beta} t},
~
~
~~~
 \th_a(t)  = \sum_{r = - \infty}^{\infty} 
 \th_{a,r} e^{\f{2 \pi i r }{\beta} t}
\ee
where, $n$ is an integer and $r$ is a half-integer.

\noindent
So, in terms of these modes, the action reduces to

\br  
S &=& \f{i \beta}{4 g l } 
\left\{ - \sum_{n= - \infty;n \neq 0}^{\infty}
\f{4 \pi^2  n^2}{\beta^2}
X^i_{a,n} X^i_{a,-n} 
\right.
+
\sum_r
 \f{4 \pi r}{\beta} 
 \th_{a,r}  \th_{a,-r}  
\non &-& \f{2 i }{l_s^2} 
 \sum_{\stackrel{r,s,l = - \infty,}{\footnotesize {\em  s+r+l = 0}}}^{\infty}
 \th_{a,r} \gamma_0 \gamma_\mu \th_{b,s}
 X^\mu_{c,l} 
\epsilon^{abc}
+
\f{ n \pi \epsilon^{abc} }{ 2 \ l_s^2 \beta }
 \sum_{ n + l + p = 0} 
X^0_{a,l} X^i_{b,m} X^i_{c,n}
\non &+& 
{1\over 2 l_s^4}
 \sum_{\stackrel{n,m,l,p = - \infty,}{ n+m+l+p=0}}^{\infty}
 X^\mu_{a,n} X^\nu_{b,m} X^\mu_{a,l}
 X^\nu_{b,p} 
- \left.
{1\over 2 l_s^4}
 \sum_{ { \stackrel{n,m,l,p = - \infty,}{n+m+l+p=0} 
 }}^{\infty}
 X^\mu_{a,n} X^\nu_{b,m} X^\mu_{b,l}
 X^\nu_{a,p}\right\} 
\label{acn6}\er  

\noindent
$n$, $m$, $l$, $p$ are integers and $r$, $s$ are half-integers.

\noindent
Here, $X^\mu_{a,n}, \beta$ have the dimension of length (L) and $\th$ has 
the dimension $L^{\f{1}{2}}$. We scale $X^\mu_{a,n}, \beta$  with 
a factor of $\f{1}{l_s}$, and $\th$ with a factor $l_s^{\f{-1}{2}}$
 to make them dimensionless, which is equivalent
to replacing all the $l_s$ in the action by 1. 

\noindent 
The first and the second terms in the action give the masses of the modes of the vector and 
the spinor fields respectively, namely, $\f{2 \pi  n}{\beta}$ and 
$ \f{2 \pi s }{\beta} $. 

\noindent
The partition function with this action is 

\bee
Z = \int e^{-iS}
\ee

\noindent
In the next section we will try to calculate this .

\section{The Partition Function and the effective Potential.}\label{partf}
\setcounter{equation}{0}

\subsection{Pfaffian}
\setcounter{equation}{0}

\noindent
The fermionic terms in the action are of the form 
$(\th_{a,r} \gamma_0\gamma_\mu  \th_{b,s} X^\mu_{c,n})_{n+r+s=0}$
and  $\th_{a,r}  \th_{b,-r}$.  In $\beta \rightarrow 0$ limit,
the first term in the action (\ref{acn6}) is dominant 
and $(X^\mu_{a,n})_{n \neq 0}$ is of the order $\sqrt{\beta}$.
So, in large temperature limit
the first term contributes to the partition function
in the leading order of $\beta$ only for $n = 0$  
$(\th_{a,r} \gamma_0\gamma_\mu  \th_{b,-r} X^\mu_{c,0})$ .
With these terms  
the action takes the form

\brr
S_f &=& \f{i\beta}{4 g}
\sum_{r=0}^\infty \left\{
 \f{4 \pi r }{\beta} 
 \th_{a, r }  \th_{a, -r }
- \f{4 \pi r}{\beta} 
 \th_{a, -r }  \th_{a, r }
- 2i
 \th_{a, r } \gamma_0 \gamma_\mu\th_{b, -r }
 X^\mu_{c,0}
\epsilon^{abc}
 - 2i
 \th_{a, -r } \gamma_0 \gamma_\mu \th_{b, r }
 X^\mu_{c,0}
\epsilon^{abc}
\right\}
\err

\noindent
Now, we will try to find out the pfaffian

\bee
Z_f = \int \prod_{a=1}^{3} d^{16}  \th_{a, r}
d^{16}  \th_{a, -r}
e^{-iS_f} 
\ee

\noindent
for this action

\br  
Z_f  &=& 
\prod_{r =0 }^\infty \int  \prod_{a=1}^{3} d^{16}  \th_{a, r}
d^{16}  \th_{a, -r}
\exp \left[
\f{-\beta}{2 g}
\left\{
\th_{a, r }
\left( \f{2 \pi r }{\beta}
 \delta^{ab}
- i 
 X^\mu_{c,0}
\epsilon^{abc} \gamma_0 \gamma_\mu
\right)  \th_{b, -r }
\right. \right.
\non&&~~~~~~~~~~~~~~~~~~~~~~~~~~~
\left. \left.
- \th_{a, -r }
\left(
 \f{2 \pi r }{\beta}
 \delta^{ab}
+ i
 X^\mu_{c,0}
\epsilon^{abc} \gamma_0 \gamma_\mu
\right) \th_{b, r }
\right\}\right]
\er  

\noindent
We rotate $X^\mu_{c,0}$ by a Lorentz transformation so that only $X_{c,0}^0$,
$X_{c,0}^1$ and $X_{c,0}^2$ are nonzero. We take the representation of the 
Gamma matrices, in which

\bee
\gamma_0 = i \sigma_2 \otimes 1_8, ~~~~~~~~~~ 
\gamma_1 =  \sigma_3 \otimes 1_8 , ~~~~~~~~~~
\gamma_2 = - \sigma_1 \otimes 1_8 
\ee

\noindent
With this choice of representation we can write

\bee
S_f =
\f{i\beta}{ g}
\sum_{r = - \infty}^{\infty} 
\left\{
\th_{a, r }
\left( \f{2 \pi r}{\beta}
\delta^{ab}
-i
\epsilon^{abc}
\left( X_c^0 1_2 + X_c^1 \sigma_1 + X_c^2 \sigma_3 \right) 
\right) \otimes 1_8       
\th_{b, -r }
\right\}
\ee 

\noindent
So, the pfaffian will be

\[
Z_f 
=
\left[
\int \prod_{a,b=1}^3 d \th'_{a, r }d  \th'_{b, -r }
\exp \left[
\f{\beta}{ g}
\left\{
\th'_{a, r }
\left( \f{2 \pi r}{\beta}
\delta^{ab}
-i
\epsilon^{abc}
\left( X_{c,0}^0 1_2 + X_{c,0}^1 \sigma_1 + X_{c,0}^2 \sigma_3 \right)
\right) 
\th'_{b, -r }
\right\}
\right]
\right]^8
\]

\noindent
where $\th'_{a, r }$ and $\th'_{b, -r }$
are the spinors in three dimension, and has two components.

\br
Z_f &=& \frac{2^{16} \pi^{48}}{g^{24}}
\prod_{r>0} \left( 16 r^6 
+\f{8 r^4 \beta^2}{\pi^2} X^\mu_{a,0}  X^\mu_{a,0}  
+ \f{\beta^4 r^2}{\pi^4} \left( ( X^\mu_{a,0}  X^\mu_{a,0} )^2 
- 4 (X^0_{a,0} X^0_{b,0} X^1_{a,0} X^1_{b,0})
\right. \right. \non &&  \left.\left.
- 4(X^0_{a,0} X^0_{b,0} X^2_{a,0} X^2_{b,0})
 \right)
+  \f{ \beta^6}{\pi^6}
\left(\epsilon_{abc} \epsilon^{\mu \nu \gamma}
X^\mu_{a,0}  X^\nu_{b,0}  X^\gamma_{c,0}\right)^2
 \right)^{8}
\er

\noindent
We can see that the above expression has  $SO(3)$ symmetry 
in spinor indices, and $SO(2,1)$ symmetry in the vector indices.
The $16 r^6$ term gives the free fermionic contribution. Note that it is 
temperature independent as the Hamiltonian is identically 
zero for free fermions in $0+1$ dimensions.

\subsection{Free Bosonic sector}
\setcounter{equation}{0}

\noindent
After doing the fermionic integral, the partition function is

\br  
Z &=& \prod_{ a = 1}^3 \int d^{10} X^\mu_{a,n} 
\left[ \frac{2^{16} \pi^{24}}{g^{24}}
\prod_r \left( 4 r^3 +  \f{ \beta^3}{\pi^3}
\epsilon_{abc} \epsilon^{\mu \nu \gamma}
X^\mu_{a,n}  X^\nu_{b,n}  X^\gamma_{c,n}
+\f{ r \beta^2}{\pi^2} X^\mu_{a,n}  X^\mu_{a,n}  \right)^{16}
\right]
\non &&
\exp \left[\f{- \beta}{4 g  }
\left\{ - \sum_{\stackrel{n= - \infty;}{n \neq 0}}^{\infty}
\f{4 \pi^2  n^2}{\beta^2}
X^i_{a,n} X^i_{a,-n}
\right.
\right.
-
\f{ n \pi \epsilon^{abc} }{ 2   \beta }
 \sum_{ n + l + p = 0}
X^0_{a,l} X^i_{b,m} X^i_{c,n}
\non &+&
{1\over 2 }
 \sum_{\stackrel{n,m,l,p = - \infty,}{ n+m+l+p=0}}^{\infty}
 X^\mu_{a,n} X^\nu_{b,m} X^\mu_{a,l}
 X^\nu_{b,p}
- \left.
 \left.
{1\over 2 }
 \sum_{\stackrel{n,m,l,p = - \infty,}{ n+m+l+p=0}}^{\infty}
 X^\mu_{a,n} X^\nu_{b,m} X^\mu_{b,l}
 X^\nu_{a,p}\right\} \right]
\er  

In infinite temperature limit $ie$ $\beta \rightarrow
0$, the first term dominates. To see the comparitive $\beta$ 
dependence of the other terms; \\
1) we set 
$ (X^\mu_{a,n} )_{n \neq 0} \rightarrow  \sqrt{\beta} X^\mu_{a,n} $\\
2) keep the terms contributing to the leading order of the partition function
in  $\beta \rightarrow 0$ limit
(this is justified in Appendix \ref{appena}).
\\
3) transform back 
$ \sqrt{\beta} X^\mu_{a,n} \rightarrow (X^\mu_{a,n} )_{n \neq 0} $\\
and we can write the partition function up to a numerical factor

\br
Z_{boson} &=& 
\f{1}{g^{24}} 
\int  d^D X^i_{a,n} d^D X^i_{a,-n}
\exp \left[ \f{\beta}{4 g}
\left\{ -\sum_{n= - \infty;n \neq 0}^{\infty}
\f{4 \pi^2  n^2}{\beta^2}
(X^i_{a,n}  X^i_{a,-n})\
\right\} \right]
\non &&
 \int d^D X_0 \prod_s d^D \th_s
\exp \left[ \f{\beta}{4 g}
\left\{
{1\over 2}
( X_{a,0}. X_{a,0} )(X_{b,0} .
 X_{b,0})
-{1\over 2}
( X_{a,0}. X_{b,0})( X_{b,0} .
 X_{a,0})
\right\} \right]
\er  

\noindent
Thus $ Z_{boson} = \int e^{-iS} = Z_{free} Z_0 $  where the first part of 
the partition function $Z_{free} $ is just the free bosonic particle
partition function (per unit volume), which is 

\[
Z_{free} =
\prod_n \prod_{a=1}^3 \prod_{i=1}^9   
\int
 d^D X^i_{a,n} d^D X^i_{a,-n}
\exp \left[ \f{\beta}{4 g}
\left\{ -\sum_{\stackrel{n= - \infty;}{n \neq 0}}^{\infty}
\f{4 \pi^2  n^2}{\beta^2}
(X^i_{a,n}  X^i_{a,-n})
\right\}\right]
= \prod_{n \neq 0} \left( \f{\sqrt{\beta g \pi}}{\pi n } \right)^{54}
\]

\noindent
Now, using $\sum_{n =1}^\infty   n^{-s} = \zeta(s)$ and $ \f{d}{ds} \zeta(s)
= - \sum_{n= 1}^\infty n^{-s} \log n $ and $\zeta(0) = 1, \zeta'(0)=
- \f{1}{2} \log (2 \pi) $ 
 (where $\zeta(s)$ is the 
Riemann Zeta function), we get 

\bee
Z_{free} = \left( \f{1}{\beta g} \right)^{27}
\ee

\noindent
Note that the free fermionic contribution contribution was discussed in the previous
 section.
$Z_0$ is calculated below.

\label{lead}\subsection{ Leading Interaction Term.}
\setcounter{equation}{0}

Now we try to calculate the 
effect of the interactions.

\noindent
As argued earlier the leading $\beta$ dependence is given by the zero modes, so
in the first approximation we drop the terms in the action involving
the higher modes. Thus we get,

\br
S_0 &=& \f{i\beta}{8 g} \left\{
 X^\mu_{a,0} X^\nu_{b,0} X^\mu_{a,0}
 X^\nu_{b,0}
- 
 X^\mu_{a,0} X^\nu_{b,0} X^\mu_{b,0}
 X^\nu_{a,0}\right\}
\non &=&
\f{i\beta}{8 g} \left\{
(X_{1,0})^2(X_{2,0})^2 +(X_{2,0})^2(X_{3,0})^2+(X_{3,0})^2(X_{1,0})^2 
\right. \non && \left.~~~~~~~~~~~~~~~~~~
- (X_{1,0}.X_{2,0})^2 - (X_{2,0}.X_{3,0})^2 - (X_{3,0}.X_{1,0})^2 
\right\} \label{ac-zero}
\label{appac}\er

\noindent
We would like to first calculate the leading order contribution to $Z$ 
that is 

\bel{par0}
Z_0 = \int dX^\mu_{a,0} e^{-i S_0}  
\ee

\noindent
Now, consider the parametrisation 

\bel{param}
X_{1,0} = ( x_1, \vec{r}_1 ), ~~~~~~
X_{2,0} = ( x_2, \vec{r}_2 ),  ~~~~~~
X_{3,0} = ( l, 0 ) 
\ee 

\noindent
If we had considered the original action with all the modes
the action would not be Lorentz invariant, for example the terms $X^0_{a,l} X^i_{b,m}X^i_{c,n}$
and $X^i_{a,n} X^i_{a,-n}$ in the original action are not Lorentz invariant. 
However, expression (\ref{ac-zero})
has Lorentz invariance, and  we are justified in using  this parametrisation
in order to evaluate (\ref{par0}).

\noindent
Under this parametrisation, the action (\ref{ac-zero}) takes the form 

\bee
S_0 = \f{i\beta}{8 g} \left\{
r_1^2  r_2^2 \sin^2 \alpha +x_1^2  r_2^2 +r_1^2 x_2^2 + r_2^2  l^2 
+l^2  r_1^2 +   
-2 x_1. x_2r_1 r_2 \cos \alpha  
\right\}
\ee

\noindent
The partition function is 

\bee
Z_0 = \int d^{10} X^\mu_{1,0} d^{10} X^\mu_{2,0}d^{10} X^\mu_{3,0}
e^{-iS_0}
\ee

\noindent
At this stage, we can find the temperature dependence of the partition function 
and the mean square separation of two D-0 branes from simple scaling argument. 
 As we have seen the leading order
$i.e.$ the zero mode contribution of the partition function comes from the term
 ${1\over 2 l_s^4}[X^\mu,X^\nu]^2 $ in the Lagrangian (\ref{ac-zero})
. And the above 
parametrisation
we have used here is also valid for  $SU(N)$ matrix model,
only the functional form of ${1\over 2 l_s^4}[X^\mu,X^\nu]^2 $
in terms of this parametrisation will be different for different $N$, but
in each case the function will be homogeneous in
$l, r_i, x_i$ where $0 < i < N^2 -2$  and of order 4. So, in each case
we need to scale these variables by  $\beta^{-\f{1}{4}} g^{\f{1}{4}} $
to scale out the $\beta$ from the exponent. 
And the temperature dependence of $\langle l^2 \rangle$
will be $\beta^{-\f{1}{2}} g^{\f{1}{2}}$ in the leading order.
Under the scaling above the measure in $Z_0$  will pick up
a $\beta^{-\f{15}{2}} g^{\f{15}{2}}$ factor for $SU(2)$
, which comes from ($3 \times 10$) 
$X^\mu_{a.n}$.  In general for $SU(N)$ in $D$ dimension there will be 
$ D (N^2 - 1)$ $X^\mu_{a,n}$ in the measure. So, the partition function $Z_0$ 
has temperature dependence  $\beta^{-\f{D (N^2 - 1)}{4}} g^{\f{D (N^2 - 1)}{4}}$.
And $Z_{free}$ will be proportional to $ (\beta g)^{(D-1)(N^2 -1)}$.

\noindent
Now we evaluate the partition function for this action.
 
\br
Z_0 &=& 
\int dl d \Omega^{(9)} l^9 \int d x_1 d x_2 \int d r_1 d r_2 d \Omega_1^{(8)}
 d \Omega_2^{(7)}
d \alpha r_1^8 r_2^8 \sin^7 \alpha 
e^{-iS}
\er

\noindent
Using
$ \int d \Omega^{(n)} = 2^{(n-1)} \pi $,
and after integration over $x_1$  , $x_2$ and $r_2$ the partition function
 is

\br
Z_0 &=&
\f{2^{37} g^5 \pi^4 \Gamma(4)}{\beta^5}
\int_{-\infty}^\infty dl l^9
\int_0^\infty d r_1 r_1^7
\exp \left[
- \f{\beta}{8 g}\left\{ l^2  r_1^2\right\} \right]
\non &&~~~~
\int_0^{\pi} d \alpha \sin^6 \alpha
\left\{ (r_1^2+l^2)  \sin^2 \alpha + l^2 \cos^2 \alpha \right\}^{-4}
\er

\noindent
Integrating over $\alpha$  \cite{GRAD}  the partition
function can be rewritten as,

\br
Z_0 &=&
\f{2^{34} 15 ~g^5 \pi^5 }{\beta^5}
\int_{-\infty}^\infty dl l^8
\int_0^\infty 
\f{d r_1 r_1^7}{(r_1^2+l^2)^{7/2}}
\exp \left[
- \left(\f{\beta l^2}{8 g}\right)  r_1^2 \right]
\er

\noindent
If we scale $r_1$ and $l$ by a factor $\beta^{-\f{1}{4}} g^{\f{1}{4}}$, the integral 
reduces to 

\br
Z_0 &=& 
\f{2^{34} 15 ~g^{\f{15}{2}} \pi^5 }{\beta^{\f{15}{2}}}
\int_{-\infty}^\infty dl l^8
\int_0^\infty
\f{d r_1 r_1^7}{(r_1^2+l^2)^{7/2}}
\exp \left[
- \left(\f{l^2}{8}\right)  r_1^2 \right]
\er

\noindent
The integral over $r_1$ can be done to give 

\br
Z_0 &=&
\f{2^{34} 15 ~g^{\f{15}{2}} \pi^5 }{\beta^{\f{15}{2}}}
\int_{-\infty}^\infty dl l^8
\exp \left(l^4/8 \right)
\left\{
 a^{-1}
\Gamma\left(\f{1}{2}, (l^4/8 ) \right)
- 3 l^2 a
\Gamma\left(\f{-1}{2}, (l^4/8) \right)
\right. \non && ~~~~~~~~~\left.
+ 3 l^4 a^{3}
\Gamma\left(\f{-3}{2}, (l^4/8) \right)
- l^6  a^{5}
\Gamma\left(\f{-5}{2}, (l^4/8) \right)
\right\}
\label{lpartf}\er

\noindent
where $\Gamma( \alpha , x ) $ is the incomplete Gamma function.

\noindent
In large $l$ regime using the asymptotic expression for the incomplete 
Gamma function 
\cite{GRAD}
we can write this expression as

\brr
Z_0 &=&
\f{2^{34} 15 ~g^{\f{15}{2}} \pi^5 }{\beta^{\f{15}{2}}}
\int_{-\infty}^\infty dl l^8
e^{-\left(\f{ l^4}{8} \right)}
\left[ \sum_{m=0}^{M-1}
(-1)^m 8^{( \f{m}{2} +1 )} l^{( - 2 m - 3)}
\right. \non && \left.
\left(
\f{\Gamma(\f{1}{2} +m)}{\Gamma(\f{1}{2})}
- 3 \f{\Gamma(\f{3}{2} +m)}{\Gamma(\f{3}{2})}
+ 3 \f{\Gamma(\f{5}{2} +m)}{\Gamma(\f{5}{2})}
- \f{\Gamma(\f{7}{2} +m)}{\Gamma(\f{7}{2})}
\right)
+ O\left(|(l^4/8)|^{-M}\right) \right]
\err

\noindent
In large $l$ approximation this boils down to

\br
Z_0 &=&
\f{2^{34} 15 ~g^{\f{15}{2}} \pi^5 }{\beta^{\f{15}{2}}}
\int_{-\infty}^\infty dl l^8
\left( 24576 l^{-15} - 2752512 l^{-19} + .... \right)
\label{largelz}
\er

\noindent 
and in the small $l$ regime the above partition function  becomes
\cite{GRAD}

\br
Z_0 &=&
\f{2^{34} 15 ~g^{\f{15}{2}} \pi^5 }{\beta^{\f{15}{2}}}
\int_{-\infty}^\infty dl l^8
\sqrt{8} l^{-1}
\exp \left(\f{l^4}{8} \right)
\left[ \sqrt{\pi}
\left\{
1
- 24  l^4
+ 256 l^8
+ \f{5120}{3} l^{12}
\right\} \right.
\non &&
+
\left\{ \left.
12 \sum_{n=0}^\infty
\f{(-1)^n \sqrt{8}   l^{( 2 + 2n)}}{n!}
\left(  \f{8}{(4 n^2 -8 n -5)(4 n^2 -8 n +3 )}
\right)
\right\}
\right]
\er

\noindent
and $l \rightarrow 0$ limit  gives

\br
Z_0 &=&
\f{2^{34} 15 ~g^{\f{15}{2}} \pi^5 }{\beta^{\f{15}{2}}}
\int_{-\infty}^\infty dl l^8
\left[  \sqrt{8 \pi } l^{-1}
- \f{256}{5} l
+ \left( \sqrt{\f{\pi}{8} } - \f{256}{3} \right) l^3
+.... \right]
\label{smalllz}
\er

\noindent
The integrand 
$ l^8 \int_0^\infty \f{d r_1 r_1^7}{(r_1^2+l^2)^{7/2}}
\exp \left[ - \left(\f{l^2}{8}\right)  r_1^2 \right] $
has $l^{-7}$ dependence for large $l$ and $l^{7}$ dependence for small $l$.
Hence, it converges for both large $l$ and small $l$ and 
the integral is non-singular and independent of $\beta$. 
\noindent
So, $Z_0 $ has a temperature dependence of $T^{\f{15}{2}}$.

\subsection{Non-leading Interaction Term.}
\setcounter{equation}{0}

\noindent
In the previous section we have seen that the leading order fermionic contribution 
comes from the free fermionic terms. Here we will try to estimate the $\beta$ dependence 
of the non-leading fermionic contributions. 

\noindent
In terms of the parametrisation  in eqns. (\ref{param}), 

\bee
Z_f = \frac{2^{16} \pi^{24}}{g^{24}}
\prod_r \left( 4 r^3 +  \f{2 \beta^3}{\pi^3}
r_1 r_2 l \sin \alpha
+\f{ r \beta^2}{\pi^2} 
\left( {r_1}^2 + {r_2}^2 + (x_1)^2 + (x_2)^2 
+  l^2 \right) 
\right)^{16}
\ee

\noindent
which, in $\beta \rightarrow 0 $ can be written as 

\br
Z_f &=& g^{-24}
\prod_r \left( r^3 O(\beta^0 g^0) + O( \beta^{\f{3}{2}} g^{\f{1}{2}} )
+ r O( \beta^{\f{9}{4}} g^{\f{3}{4}} ) +.............\right)
\non &=&
g^{-24}
\left( O(1) + O( \beta^{\f{3}{2}} g^{\f{1}{2}} )
+ O(\beta^{\f{9}{4}} g^{\f{1}{4}} )
+ r O( \beta^{\f{9}{4}} g^{\f{3}{4}} ) +.............\right)
\er

\noindent
where the first term is the leading order fermionic partition function 
we have discussed  in subsection (3.2).


\noindent
So we can see that in the partition function at high temperature the contribution
of the zero modes (bosonic) is dominant. We have earlier argued that
the
 higher 
modes of the bosonic fields will also contribute in non-leading
terms.

\subsection{Mean-square Separation of the D-0 branes.}
\setcounter{equation}{0}

 As we are working in Euclidean
metric and  since for zero mode calculation we have Lorentz symmetry, we
can 
identify
$l$ as one of the spatial components and hence as the separation between 
two D-0 branes.

\noindent 
Now, we try to see the temperature dependence of the expectation value of
 $l^{2n}$

\bee
\langle l^{2n} \rangle  = \f{\int e^{-iS} l^{2n} }{Z}
\ee

\noindent
$i.e.$

\br
\langle l^{2n} \rangle
 &=&
(\beta^{-\f{1}{4}} g^{\f{1}{4}} )^{2n}
\f{
\int_{-\infty}^\infty dl l^8 l^{(2n)}
\int_0^\infty
\f{d r_1 r_1^7}{(r_1^2+l^2)^{7/2}}
\exp \left[
- \left(\f{l^2}{8 }\right)  r_1^2 \right]
}{
\int_{-\infty}^\infty dl l^8
\int_0^\infty
\f{d r_1 r_1^7}{(r_1^2+l^2)^{7/2}}
\exp \left[
- \left(\f{l^2}{8 }\right)  r_1^2 \right] }
\er

\noindent
As $l$ is the separation of two D-0 branes, we get the mean square separation of two D-0 branes from
this,  by
putting $n=1$ and doing the integral numerically, and restoring $l_s$, we get	

\bel{msq}
 \left\langle \left(\f{l}{l_s}\right)^2 \right\rangle = 6.385 \left(\f{\beta}{ g l_s}\right)^{-\f{1}{2}}  \ee

\noindent
If we assume high temperature expression has a finite radius of convergence,
we can conclude that 
the mean square separation is finite for finite temperature. This implies that there is a confining
 potential that binds the D-0 branes. As argued earlier the scaling argument that gives the $\beta$ and $g$ dependence in \ref{msq} is valid for {\em all N }. So we can 
conclude that $ \langle l^2 \rangle \approx \sqrt{\f{g}{\beta}} $ for all $N$.   

\subsection{Effective Potential.}
\setcounter{equation}{0}

For high temperature we have evaluated the partition function both for large 
and small $l$ (eqn.\ref{largelz},\ref{smalllz}).  
Up to leading order the effective 
potential between two D-0 branes is proportional 
to $- \log l$ and $ \log l$ for small and large $l$.  We can see that
the 
potential 
increases at both $l$ ends,  though we can not clearly see the nature of the 
potential in the intermediate region but we can 
conclude that the potential is a confining potential and binds the 
 D-0 branes.

\section{Conclusion}
\setcounter{equation}{0}

In this paper we have attempted to solve the $N=2$ matrix model in the
high 
temperature limit
. The leading nontrivial term of the partition function 
has been calculated exactly (eqn. \ref{lpartf}). 
 The non-leading terms can
also be 
systematically calculated 
although we haven't attempted to work them out in this paper. From a scaling argument 
we have also determined the $\beta$ and $g$ dependence of the leading term for any $N$. This 
complements the work 
of \cite{AMB}, where the one loop partition function was calculated with the 
entire $\beta$ dependence.
We find that $ \langle l^2 \rangle \propto
\sqrt{\f{g}{\beta}} $ (eqn. \ref{msq}) (true for any $N$), 
the finiteness
of which shows that there must be a potential between D-0 branes that binds
them. In \cite{BS1,AMB}  also a logarithmic and attractive potential
were found. The present calculation being exact in $g_s$ is valid at
all distances.  Thus unlike in \cite{BS1,AMB},
the (finite temperature) logarithmic potential found here is 
attractive at long distances
and repulsive at short distances so it has a minimum at non-zero
separation. Similar issue for low temperature has been discussed in \cite{MLI}. 
In \cite{BS1} it
was found that at high temperatures, 
the configuration with all the D-0-branes clustered
at the origin $i.e.$ with the zero separation,
had lower free energy than the one where they were
spread out. However, that was a large $N$ calculation and also restricted
to one loop. 
It is therefore possible that more  exact calculation will resolve this 
issue. 

As mentioned in the introduction, describing completely the dynamics of two D-0
branes in 
M-theory would
require the infinite $N$ model. Whether some high temperature
expansion of 
that model within 
the $\f{1}{N}$  approximation scheme can be attempted is an open question.

\appendix

\renewcommand{\thesection}{\Alph{section}}
\renewcommand{\theequation}{\thesection.\arabic{equation}}

\section{Comparitive $\beta$ dependence of the terms in action}\label{appena}
\setcounter{equation}{0}

We are interested in investigating the comparitive $\beta$ dependence of the 
terms in the action in eqn. (\ref{acn6}). For convenience, in this part we 
will suppress the isospin and vector indices. 
The action takes the form

\bee
S =
\f{1}{g} \left\{
- \sum_{\stackrel{n= - \infty;}{n \neq 0}}^{\infty}
\f{a_n}{\beta^2}
X_{n} X_{-n}
+\sum_{l+m+n=0} \f{f_n}{\beta} X_l X_m X_n
 -
c  \sum_{\stackrel{n,m,l,p = - \infty,}{ n+m+l+p=0}}^{\infty}
 X_{n} X_{m} X_{l}
 X_{p}
\right\}
\ee  

\noindent 
where $a_n, b_n, c$ $d$ and $f_n$ are constants and are given by 
$ a_n = \pi^2 n^2 $ ,  $c = \f{1}{8 }$, 
and $ f_n = \f{n \pi }{4}$

\noindent
Now, when we expand the sum over $n$, $m$, $l$ and $p$ in last two terms, 
we will get terms with all of these indices being $0$, with two of the indices
being $0$ and with one of them being $0$, so the action can be written in the form 
(taking one of each type, 
as the terms of the same type have same $\beta$ dependence).  

\brr
S &=& 
-\f{1}{\beta g } 
\sum_{\stackrel{n= - \infty;}{n \neq 0}}^{\infty}
a_n X_{n}^2
-
\f{c \beta}{g}  X_{0}^4
-
\f{c \beta}{g}  \sum_{n\neq 0}
 X_{0}^2 X_{n}^2
+\f{1}{g} \sum_{n \neq 0} f_n X_0 X_{-n} X_n
\non &-&
\f{ c \beta}{g}   \sum_{\stackrel{m,l,p \neq 0 ,}{ m+l+p=0}}
 X_{0} X_{m} X_{l}
 X_{p}
-
\f{ c \beta}{g}   \sum_{\stackrel{n,m,l,p \neq 0,}{ n+m+l+p=0}}
 X_{n} X_{m} X_{l}
 X_{p}
\err

\noindent
 Now, we set

\[ (X_{n} )_{n \neq 0} =  \sqrt{ g \beta } X'_{n} \]

\noindent
Under this transformation, the action reduces to

\br  
S &=& 
\left\{
- \f{c \beta}{g}  X_0^4 \right\}
+ \left\{-a_n  \sum_{n \neq 0}   {X'_n}^2 
- c\beta^2 \sum_{n \neq 0}  {X'_n}^2  X_0^2
+\sum_{n \neq 0} f_n X_0 X_{-n} X_n
\right. \non && \left.
- c\beta^{\f{5}{2}} \sqrt{g} \sum_{n+m+l=0}  X'_n X'_m X'_l X_0
- c\beta^3 g \sum_{n+m+l+p=0 } X'_n X'_m X'_l X'_p \right\}
\label{ac2} \er

\noindent
Again, we can neglect 
$\beta^{\f{5}{2}}  X'_n X'_m X'_l X_0$ and 
$\beta^3 X'_n X'_m X'_l X'_p $ terms also, as 
they have higher order $\beta$ dependence ( $\f{5}{2}$ 
and 3 respectively).
To justify this step, note that since the potential is always greater than zero, 
it is equivalent to an integral of the form

\[
Z = \int dx \exp \left(-a x^2 + \f{\beta^{\f{3}{2}} d}{\sqrt{g}}  x - \beta^2 c x^3 
- c g \beta^3 x^4 \right) 
 = \int dx e^{-a x^2} \exp \left(\f{\beta^{\f{3}{2}} d}{\sqrt{g}}  x 
- \beta^2 c x^3 - c g \beta^3 x^4 \right)
\]

\noindent
Being a  convergent integral, as $\beta \rightarrow 0 $,  we can use Taylor expansion
and write as

\bee
Z = \int e^{-a x^2} \left( 1 - \f{ \beta^{\f{3}{2}} d}{\sqrt{g} } x  
+ \beta^2 c x^3 - \f{\beta^2 d}{4 \sqrt{g}} x 
+ .... \right) dx 
\ee

\noindent
Neglecting the higher order $\beta$ terms, we can write the partition 
function for action (\ref{ac2}), 

\[
Z = 
\int dX_0 \exp \left\{ 
-\f{c \beta}{g} X_0^4 \right\} 
\prod_{n \neq 0} \int d X_n 
\exp \left\{ (- a_n + \beta f_n X_0    
- c\beta^2 X_0^2) \f{1}{g \beta } X_n^2
\right\} 
\]

\noindent
In order to justify dropping the $X_0^2 X_n^2$ and  $X_0 X_n^2$ terms, we can proceed as follows. 
After doing the $X_n$ integral the partition function reduces to 

\bee
Z =
\sqrt{\pi g \beta} \prod_{n \neq 0}
\int dX_0 
\f{\exp \left\{ 
-\f{c \beta}{g} X_0^4 
\right\}}{( a_n -f_n \beta X_0 + c \beta^2 X_0^2)^{\f{1}{2} }}
\ee

\noindent
As here  $ 4 ( a_n -1) c - f_n^2 $ is positive, $(a_n-1) -f_n X_0 + c  X_0^2$
is also positive , $i.e.$  \\  $[a_n-f_nX_0+cX_0^2]>1$. So

\bee
\f{\exp \left\{
-\f{c \beta}{g} X_0^4
\right\}}{( a_n -f_n \beta X_0 + c \beta^2 X_0^2)^{\f{1}{2} }}
< \exp \left\{
-\f{c \beta}{g} X_0^4
\right\}
\ee

\noindent
and the integral is finite. So in $\beta \rightarrow 0 $ limit the leading order term can be given  
by $\int dX_0
a_n^{-\f{1}{2}} \exp \left\{
-\f{c \beta}{g} X_0^4
\right\} $
and 
the corrections will  vanish at the positive power of $\beta$ .
So, in $\beta \rightarrow 0$ limit we can ignore those,
which is equivalent to ignoring the $f_n \beta X_0 X_n X_m $ and $c \beta^2
X_0^2 X_n^2$ terms in action so it can be written  as     

\bee
S =
\f{a}{\beta}
\sum_{n= - \infty;n \neq 0}^{\infty}
X_{n}^2
-
c \beta X_{0}^4
\ee

\noindent
which is also eqn.(\ref{appac}).

\begin{center}
{\bf Acknowledgements}
\end{center}

We thank N.D. Hari Dass,
S. Kalyana Rama and P. Majumdar for useful discussions and suggestions.
Subrata Bal likes to thank T. Sarkar, A. Dasgupta and S. Murugesh for some 
discussions.

\end{document}